# A revised comparison between FF five-factor model and three-factor model——based on China's A-share market

Zhijing Zhang[a], Yue Yu[b], Qinghua Ma[b,1], Haixiang Yao[c,d]

[a] Youshi Limited Company (China) of Science and Technology, Guangzhou 510335, China

[b] School of Mathematics and Statistics, Guangdong University of Foreign Studies, Guangzhou 510006, China

[c] School of Finance, Guangdong University of Foreign Studies, Guangzhou 510006, China

[d] Southern China Institute of Fortune Management Research (IFMR), Guangzhou 51006, China

**Abstract:** In allusion to some contradicting results in existing research, this paper selects China's latest stock data from 2005 to 2020 for empirical analysis. By choosing this periods' data, we avoid the periods of China's significant stock market reforms to reduce the impact of the government's policy on the factor effect. In this paper, the redundant factors (HML, CMA) are orthogonalized, and the regression analysis of 5*5 portfolio of Size-B/M and Size-Inv is carried out with these two orthogonalized factors. It found that the HML and the CMA are still significant in many portfolios, indicating that they have a strong explanatory ability, which is also consistent with the results of GRS test. All these show that the five-factor model has a better ability to explain the excess return rate. In the concrete analysis, this paper uses the methods of the five-factor 25-group portfolio returns calculation, the five-factor regression analysis, the orthogonal treatment, the five-factor 25-group regression and the GRS test to more comprehensively explain the excellent explanatory ability of the five-factor model to the excess return. Then, we analyze the possible reasons for the strong explanatory ability of the HML, CMA and RMW from the aspects of price to book ratio, turnover rate and correlation coefficient. We also give a detailed explanation of the results, and analyze the changes of China's stock market policy and investors' investment style recent years. Finally, this paper attempts to put forward some useful suggestions on the development of asset pricing model and China's stock market.

**Key words**: Finance, Asset pricing, Empirical research, A-share market, Five-factor model

## 1. Introduction

Asset pricing is one of the most important major topics in modern finance. Its main value is to quantitatively analysis the factors that affect the expected return of different portfolios in the capital market. The classical capital asset pricing model (CAPM) was put forward by economists and financiers such as Sharpe (1964), Lintner (1965) and Mossin (1966) based on the data of the American stock market. The main content of the theory is that there is a positive correlation between the expectation of the excess return of a portfolio (compared with the risk-free assets) and the measure of asset risk

[1] Corresponding author.
   *E-mail addresses:* 18027194376@163.com (Z. Zhang); yuyue1260@foxmail.com (Y. Yu);
   mqh@gdufs.edu.cn (Q. Ma); yaohaixiang@gdufs.edu.cn (X. Yao)

"β". With the development of the CAPM model, Markowitz's portfolio selection theory has made a breakthrough in its application, and securities theory has changed from simple qualitative analysis to quantitative analysis, from theoretical explanation to empirical verification, which has become the pillar of modern financial market price theory.

Although CAPM has greatly promoted the progress in the field of asset pricing, as a single factor model which uses a single risk factor to describe the expected rate of return of assets, it has some problems such as some idealistic assumptions (it is difficult to realize perfect competition market, friction and information asymmetry in trading market, etc.), the difficulty of how to determine β value. It also faces many challenges from empirical test. For example, Fan Longzhen et al. (2004) analyzed that the size effect and value effect are obvious in China's Shanghai and Shenzhen markets, and these effects cannot match the explanation of β Value in the CAPM.

With the continuous popularization and application of CAPM model, more and more scholars have found that the factors such as company size and B/M can better explain the expected return of stock. In this context, Fama and French, using the stock return data of American stock exchanges in the 1970s and 1980s, found two factors—Size and B/M which enjoy better explanation the change of excess returns than the "β" in the CAPM. Fama and French proposed that these two factors are exactly the other part of the influencing factors that the β Value in CAPM model cannot reflect. After their continuous in-depth research, Fama and French published their classic paper in the field of asset pricing in 1993 "Common Risk Factors in the Returns on Stocks and Bonds ". In this paper, they proposed a new capital asset pricing model with Size factor (SMB) and value factor (HML) – the three-factor model. The breakthrough of this model is that the size factor and value factor are not the market value or B/M itself, but the income difference of different portfolios. This construction mode has become a common method to construct factors in the financial field after that.

Following the publication of this paper, scholars from all over the world have also made many in-depth studies on the analytical capability of the three-factor model and the difference between it and the CAPM model. Griffin (2002) conducted an in-depth empirical test on the three-factor model based on the securities market data of Canada, Japan and the United Kingdom and found that the three-factor model has better analytical power on the difference of stock portfolio returns than the CAPM model. Cao et al. (2005) compared the analytical capability of the three-factor model and the CAPM model in the China's market and found that the former has better analytical ability. It can conclude that the three-factor model has been empirically tested for many times in different securities markets, and the results are consistent in the academic and investment areas: the three-factor model has better analytical ability than the CAPM model.

At the same time, scholars found that for some financial anomalies and the three-factor model is difficult to show enough explanatory power. Novy and Marx (2013) found that expect for the three factors in the model, there is also a "profit expectation factor" which has a significant correlation with the expected value of stock return. Aharoni et al. (2013) also found that there is a significant correlation between the investment situation of different enterprises and the expected value of the excess returns of the enterprise's stock; Fama-French (2006, 2008) also confirmed that there is a significant correlation between investment factors and stock return expectation.

Inspired by the above and not mentioned here research and the Dividend Discount Model proposed by John Burr Williams, Fama-French (2015) added the profit factor (RMW) representing the company's profitability and the investment factor (CMA) representing the company's investment status, combined with the market factor (MKT), the size factor (SMB) and value factor (HML) in the three-factor model, to propose the Fama-French five-factor model. They also used the data of American and European stock markets to verify the new model's better analytic ability.

We find that the performance of the five-factor model is different in different regions and countries due to different policies, mechanisms and development degrees of stock markets. Fama and French (2016) found that the U.S. stock market has significant momentum effect, but the five-factor model cannot well explain the momentum effect. Adding momentum factor can significantly improve the performance of this model. James Foye (2018) discussed the application of the five-factor model in three different regions (18 countries), pointing out that the five-factor model is better than the three-factor model in Eastern Europe and Latin America, but not as good as the three-factor model in Asian markets (emerging markets). Shaun Cox and James Britten (2019) discussed the performance of the five-factor model in Johannesburg Stock Exchange (JSE), considered that it has the similar explanatory ability as the three-factor size-value model, and concluded that profitability is more significant than investment and the asset pricing models on the JSE will get improved because of the addition of profitability. Philipp dirkx and Franziska J. Peter (2020) used the six-factor model with momentum factor to analyze the return premium in the Germany's market, and concluded that the newly added probability and investment factors do not have excellent explanatory ability, and the performance of the six-factor model is not as good as the three-factor model. James Foye and Aljoša Valentinčič (2020) pointed out that Indonesia has not established a mature financial reporting mechanism, resulting in that the accounting-based factors (probability and investment) cannot well explain the average return of the portfolios. M. N. López-García (2021) added MOM momentum factor and new memory factor H to form a new five-factor model, and pointed that H is more significant than MOM. The above all also inspires our interest in analyzing the explanatory power of the five-factor model in the China's stock market.

Despite the fact that it has been six years since the five-factor model was put forward, scholars still doubt the applicability and analytical ability of the five-factor model in China's stock market due to its issues such as information asymmetry which exists less in more mature markets. At present, there are only a few high-quality studies on the applicability of the RMW and the CMA in China's trading market and scholars also hold different attitudes on how to improve the five-factor model. Zhao Shengmin et al. (2016) found that the size effect and value effect of China's stock market are obvious, while the OP and Inv are not helpful to explain the return of stock portfolio, and the three-factors model are more suitable for China's investment market. Guo et al. (2017) pointed out that in China's A-share stock market, the three factors: size, value and profitability, have relatively strong explanatory power to the stock excess return, while in contrast, the investment factor is redundant. Li Zhibing et al. (2017) stated that the five-factor model is more suitable than the CAMP, the three-factor model and the Carhart four-factor model in China's A-share market. By testing the Chinese stock market in different periods, they point out that the significance of factors is different in different periods. Wenting jiao et al. (2017) pointed that the five-factor model does not capture more changes of average excess stock return than the three-factor model based on the data of China's stock market from 2010 to 2015. They also used U.S. data for regression in the same period. The results show that five-factor model performes slightly better in US stock market than that in China's A-share market. Jianan Liu et al. (2019) pointed that the returns of China's small listed stocks are affected by China's tight IPO policies, and the performance of the factors formed by eliminating the smallest 30% of stocks is much better than that formed by using all stocks.

We can see that the applicability of the five-factor model in different stock markets is quite different. As far as China's A-share stock market is concerned, whether the analytical ability of the five-factor model is better than that of the three-factor one has not yet been determined.

In this context, our paper hopes to conduct an empirical test on the applicability of the five-factor model by using the latest trading data and enterprise data of China's A-share market, and compare the differences between the analytical capacity of the five-factor model and the three-factor model, so as to provide supplement for the research of the factor model in asset pricing field, and evidence for the development of China's capital asset pricing field.

## 2. Description of the Five-Factor Model

The five-factor model is based on the three-factor model. It refers to Williams' theory of further splitting stock dividend in 1938's "Dividend Discount Model" (DDM), adding RMW and CMA to the three-factor model, to form a five-factor model. With the application of dividend discount model, Fama and French's asset pricing model is upgraded from the simple three-factor model based on market data regression to the five-factor model based on certain economic theory model.

The basic form of dividend discount model is as follows:

$$V_t = \sum_{t=1}^{\infty} \frac{D_{t+\tau}}{(1+k)^\tau}$$

Where $V_t$ represents the internal value of the stock, $D_{t+\tau}$ represents the expected value of the dividend amount in t+τ period, and k represents the expected value of the long-term yield of the stock.

It can be seen from the equation that if two stocks in the same period have the same expected dividend $D_t$, the stocks with high intrinsic value will be considered to enjoy lower average expected return. On this basis, we can divide the dividend per share into the relationship between the expected profit and the expected investment:

$$D_{t+\tau} = Y_{t+\tau} - dB_{t+\tau}$$

Where $Y_{t+\tau}$ is the equity income of the stock from t to t+τ, representing the profitability of the enterprise; $dB_{t+\tau}$ is the increment of owner's equity in the current period, representing the investment situation of the enterprise. If the owner's right at the beginning of the period is written as $B_t$, the dividend discount model can be expressed as follows:

$$\frac{V_t}{B_t} = \sum_{\tau=1}^{\infty} \frac{Y_{t+\tau} - dB_{t+\tau}}{(1+k)^\tau \times B_t}$$

From this formula, we can analyze that the expected return rate k of the stock or portfolio is affected by the current intrinsic value $V_t$ and the future return $D_{t+\tau}$. If the future return flow is further decomposed, we can get that the expected return rate of the stock or portfolio is related to the expected profit $Y_{t+\tau}$ of the enterprise and the expected investment $dB_{t+\tau}$. Therefore, the higher the expected profit $Y_t$, the higher the expected return k is; with the increase of investment input, the expected value k of stock return increases in the opposite direction. This is the reason why Fama and French added the profit factor and investment factor to the three-factor model as the explanatory factors of stock excess return, and further put forward the theoretical basis of Fama-French's five factor model.

After adding the RMW and the CMA, combined with the SMB and the HML in the three-factor model, we can put forward the final form of Fama-French's five factor model:

$$R_{it} - R_{Ft} = a_i + b_i(R_{Mt} - R_{Ft}) + s_i SMB_t + h_i HML_t + r_i RMW_t + c_i CMW_t + e_{it}$$

Among them, $R_{it} - R_{Ft}$ represents the excess return of stock i in t period; $R_{Mt} - R_{Ft}$ is the excess return of market portfolio weighted by market value which is used to measure the premium from market risk in the five-factor model.

SMB (small minus big) represents the difference of the return rate between the small market value company's stock portfolios and the large market value company's; HML (high minus low) represents the difference of the return rate between the stock portfolios of the company with high B/M and the stock portfolios of the company with low B/M; RMW (robust minus weak) represents the difference of return rate of high profit company's stock portfolio and low profit company's portfolio, and CMA (conservative minus aggressive) represents the difference of return rate of low investment company's stock portfolio and high investment company's; $e_{it}$ is the regression residual with an expected value of 0.

# 3. Empirical Analysis

### (1) Data Description

In this section, we select the A-share data of Shanghai and Shenzhen stock markets from January 1, 2005 to December 31, 2020 as the analysis sample excluding the stock data of ST (Special Treatment) and PT (Particular Transfer). The data we use comes from China Stock Market & Accounting Research Database (CSMAR). We mainly use the following data set: the yield of every stock in each period, closing price, the risk-free interest, the balance sheet, income statement and so on. Based on these real data, we construct the five-factor model.

### (2) Empirical Test of Factor Effect

According to the method proposed by Fama-French in the five-factor model, the entire stock market is divided into five groups respectively according to the Size, B/M, OP and Inv. At the same time, the factors of Size, B/M, OP and Inv are combined with each other to form portfolios: Size and B/M, Size and OP, Size and Inv. Each portfolio includes 25 combinations generated by two groups crossing, and then calculates the average monthly excess returns of each portfolio.

The table below shows the average monthly excess returns of the portfolios formed in the above way:

Table 4.2 Size-B/M portfolios

| Size-B/M | Low | 2 | 3 | 4 | High |
|---|---|---|---|---|---|
| Small | 1.851196 | 2.128478 | 2.148923 | 2.151535 | 1.943599 |
| 2 | 1.476902 | 1.786545 | 1.913537 | 2.036226 | 1.727048 |
| 3 | 1.376502 | 1.724378 | 1.61121 | 1.660342 | 1.624684 |
| 4 | 1.355968 | 1.504806 | 1.626458 | 1.74522 | 1.408737 |
| Big | 1.62876 | 1.18855 | 1.241463 | 1.314614 | 1.049259 |

Table 4.3 Size-OP portfolios

| Size-OP | Low | 2 | 3 | 4 | High |
|---|---|---|---|---|---|
| Small | 2.06888 | 1.924582 | 2.165042 | 2.081513 | 1.855556 |
| 2 | 1.724693 | 1.821312 | 1.885216 | 1.922984 | 1.756523 |
| 3 | 1.455961 | 1.539619 | 1.581122 | 1.760784 | 1.718851 |
| 4 | 1.235418 | 1.491537 | 1.468308 | 1.575377 | 1.703043 |
| Big | 1.16597 | 1.02307 | 0.9434643 | 1.120409 | 1.627726 |

Table 4.4 Size-Inv portfolios

| Size-Inv | Low | 2 | 3 | 4 | High |
|---|---|---|---|---|---|
| Small | 1.957998 | 2.064537 | 2.009726 | 2.221085 | 1.971111 |
| 2 | 1.75322 | 1.898721 | 1.873128 | 1.812493 | 1.700221 |
| 3 | 1.500576 | 1.570828 | 1.594425 | 1.558031 | 1.76931 |
| 4 | 1.495644 | 1.46229 | 1.580673 | 1.621097 | 1.438504 |
| Big | 0.9445265 | 1.090302 | 1.381016 | 1.497715 | 1.340053 |

From the perspective of the variety of portfolio yield and according to different factors, we can get the following conclusions:

Firstly, there is a very obvious size effect in China's A-share market. In each group, we can see that the big-size portfolio shows a lower average monthly excess return than the small-size one. This also verifies that as a common factor of the three-factor model and the five-factor model, the Size factor has a significant impact on stock value.

Secondly, China's A-share market also demonstrates the market value effect. From the table 4.2, we may find that the portfolio with higher B/M tends to enjoy higher excess return compared to the lower one in the same size group of stocks. In addition to the highest group of B/M, namely the fifth column of table 4.2 which shows a certain deviation (showing a tail-up phenomenon which was also found by Tian Lihui (2014) in their research on three-factor model), most of the other groups present a positive correlation between the B/M and the stock excess return. This result is consistent with the research results of scholars on the impact of B/M to the stock excess return. The main explanations are as follows: companies with high B/M usually manifest poor fundamentals and investors would irrationally underestimate the stock value of such companies. When this underestimation is corrected, the stock of enterprises with high B/M will have higher excess returns than those with low B/M.

Thirdly, in the portfolio formed on Size-OP and Size-Inv, the average monthly excess returns do not show a significant correlation with the profitability or investment, but there are still some interesting phenomena: from table 4.3 Size-OP portfolios, we can find that the size ranking in the middle three rows, namely the second, the third and the fourth row of the table 4.3, all have obvious gradual increase characteristics composed of four values. These three rows show that the more profitable the portfolios of listed companies are, the more likely they are to get higher average returns, which is close to the characteristics of the US stock market. However, in the biggest and smallest size groups, there is no obvious positive correlation between profitability and yield, namely no profitability effect. Overall, the more profitable the portfolios of the listed companies in China are, the more likely they are to get higher average returns. Besides, from table 4.4 Size-Inv portfolios, in the smallest and biggest Size quintiles, namely in the first and fifth rows of the table, there are both four values respectively, indicating that the more aggressive the investment, the greater the monthly average return of the portfolio is. There are some fluctuations in the middle three rows, but on the whole, the stronger the investment ability of the portfolios of the listed companies, the more likely they are to get higher average returns, which is contrary to the performance of the U.S. stock market. The reason may be that China's listed companies are in the stage of rapid expansion and need external debt financing to meet the needs of expansion, resulting in the increase of investment growth and the greater risk the company faces. Therefore, the higher the rate of return required to hold the company's shares.

Therefore, in the empirical test of this part, we draw a preliminary conclusion that there is an obvious size effect and a value effect in China's A-share market, and the companies' profitability and investment effect have no significant impact on the excess returns of the enterprise's stock, with the possibility to be redundant, but there are certain characteristic effects related to profitability and investment.

**(3) Factor Regression Coefficient Test of Five-factor Model**
  (3.1) The Construction of Factors
(I) MKT: $R_{Mt} - R_{Ft}$
The MKT is equal to the Market Profitability $R_{Mt}$ minus risk-free rate $R_{Ft}$. In this paper, the risk-free rate $R_{Ft}$ indicates the monthly risk-free rate derived from the CSMAR database. Market Profitability $R_{Mt}$ is the market portfolio yield with weighted average market value.
(II) SMB
Referring to the definition method of Fama-French's five-factor model, this paper will use the method of 2x3 to classify the A-shares in Shanghai and Shenzhen stock markets according to the Size, B/M, OP and Investment: The Size will be divided into big size group (B) and small size group (S) according to the median; B/M, OP and Investment dimension are divided into three groups according to the 30% and 70% quantiles. Using the big and small Size groups and the B/M groups, the OP groups and the Inv groups to conduct 2x3 cross combinations respectively, six Size-B/M combinations, six Size-OP combinations and six Size-Inv combinations can be obtained.

Table 4.5 Factor definitions using 2x3 sorts

|  | HML | RMW | CMA |
|---|---|---|---|
| Small (S) | SH,SN,SL | SR,SN,SW | SC,SN,SA |
| Big (B) | BH,BN,BL | BR,BN,BW | BC,BN,BA |

Finally, 18 combinations from three categories are averaged. The average of the three size factors (the SMB used for regression) is:

$$SMB_{B/M} = \frac{SH + SN + SL}{3} - \frac{BH + BN + BL}{3}$$

$$SMB_{OP} = \frac{SR + SN + SW}{3} - \frac{BR + BN + BW}{3}$$

$$SMB_{Inv} = \frac{SC + SN + SA}{3} - \frac{BC + BN + BA}{3}$$

$$SMB = \frac{SMB_{B/M} + SMB_{OP} + SMB_{Inv}}{3}$$

(III) HML RMW CMA

Taking the HML as an example, the value of the HML is equal to the average of the returns of the small-Size, high-B/M portfolio and the big-Size, high-B/M portfolio, minus the average of the returns of the small-Size, low-B/M portfolio and the big-Size, low-B/M portfolio. Similarly, the RMW is equal to the difference between the average return of small/big-Size, high-OP portfolio and the average return of small/big-Size, low-OP portfolio; The CMA is equal to the difference between the average return of small/big-Size, high-Inv portfolio and the average return of small/big-Size, low-Inv portfolio.

Table 4.6 The 2x3 construction of HML, RMW, CMA

| Factor | Calculation formula of each factor |
|---|---|
| HML | HML=(SH+BH)/2-(SL+BL)/2 |
| RMW | RMW=(SR+BR)/2-(SW+BW)/2 |
| CMA | CMA=(SC+BC)/2-(SA+BA)/2 |

(3.2) Regression analysis

Mainly using "bys" and "collapse" functions in STATA, we can get the monthly excess return and the value of the MKT, the HML, the RMW and the CMA in 2005-2020. Finally, STATA is used for OLS regression and the result is as follows:

| Source | SS | df | MS | | | |
|---|---|---|---|---|---|---|
| Model | 18482.4136 | 5 | 3696.48272 | Number of obs | = | 192 |
| Residual | 132.047988 | 186 | .709935419 | F(5, 186) | = | 5206.79 |
| | | | | Prob > F | = | 0.0000 |
| | | | | R-squared | = | 0.9929 |
| | | | | Adj R-squared | = | 0.9927 |
| Total | 18614.4616 | 191 | 97.4579142 | Root MSE | = | .84258 |

| monthly_ex~e | Coef. | Std. Err. | t | P>|t| | [95% Conf. Interval] | |
|---|---|---|---|---|---|---|
| MKT | 1.005396 | .0084042 | 119.63 | 0.000 | .9888167 | 1.021976 |
| SMB | .584511 | .0216458 | 27.00 | 0.000 | .5418082 | .6272138 |
| HML | -.0381215 | .0282702 | -1.35 | 0.179 | -.0938928 | .0176499 |
| RMW | .1688228 | .0347262 | 4.86 | 0.000 | .1003149 | .2373306 |
| CMA | .0581737 | .043601 | 1.33 | 0.184 | -.0278425 | .1441898 |
| _cons | .0721699 | .0647901 | 1.11 | 0.267 | -.0556481 | .1999878 |

Picture 4.3 Test results using Stata

From the $R^2$ and adjusted $R^2$ (coefficient of determination), the goodness of fit of this model is excellent: 99% of the change of excess returns in China's A-share market from 2005 to 2020 can be explained by the change of five-factor coefficient. Then, the P values of the MKT, the SMB and the RMW are all significantly less than 0.05, indicating these three factors have significant impacts on the excess returns of the stocks.

However, the t-statistic of HML is -1.35, and the p value that is 0.179. Therefore, we think that HML has no significant impact on the excess returns of stocks, and it may be a redundant factor. This is similar to the situation of the U.S. stock market. In Fama-French's (2015) paper, they found that the HML becomes redundant after adding RMW and CMA. But the difference between our result and Fama's is that in addition to the HML, our analysis suggests that the CMA may also be redundant. The t-statistic of the CMA is 1.33 with a P value of 0.184 which indicate the factor has no significant influence on the excess returns of stock, which is consistent with the Guo et al. (2017)'s results. Whether these two factors are redundant needs further verification.

Next, we discuss the regression results of 5*5 portfolios of the Size-B/M and the Size-Inv, and focus on the regression coefficients and significance of value and investment factors. For the treatment of redundant factors in Fama-French's five-factor model, we construct orthogonalization factors HMLO and CMAO to replace the original ones.

Table 4.8 the regression of Size-B/M

$$R_{it} - R_{Ft} = a_i + b_i(R_{Mt} - R_{Ft}) + s_i SMB_t + h_i HMLO_t + r_i RMW_t + c_i CMW_t + e_{it}$$

|       | h |  |  |  |  | t(h) |  |  |  |  |
|-------|-------|-------|-------|-------|-------|-------|-------|-------|-------|-------|
| Small | -0.60*** | -0.34*** | -0.03 | 0.02 | 0.09 | -6.97 | -3.17 | -0.34 | 0.31 | 1.45 |
| 2     | -0.47*** | -0.29*** | -0.27*** | -0.05 | 0.17** | -3.47 | -3.96 | -4.13 | -0.70 | 2.19 |
| 3     | -0.58*** | -0.34*** | -0.22*** | -0.10 | 0.26*** | -5.93 | -4.83 | -2.69 | -1.19 | 3.18 |
| 4     | -0.77*** | -0.34*** | -0.23*** | -0.11 | 0.21*** | -9.09 | -4.65 | -2.86 | -1.41 | 2.84 |
| Big   | -0.98*** | -0.43*** | 0.01 | 0.42*** | 0.47*** | -13.06 | -5.61 | 0.10 | 4.75 | 5.06 |

Table 4.9 the regression of Size-Inv

$$R_{it} - R_{Ft} = a_i + b_i(R_{Mt} - R_{Ft}) + s_i SMB_t + h_i HML_t + r_i RMW_t + c_i CMWO_t + e_{it}$$

|       | c |  |  |  |  | t(c) |  |  |  |  |
|-------|-------|-------|-------|-------|-------|-------|-------|-------|-------|-------|
| Small | 0.33*** | 0.10 | 0.04 | -0.21* | -0.53*** | 2.61 | 1.35 | 0.34 | -1.78 | -3.69 |
| 2     | 0.39*** | 0.28*** | 0.18 | -0.07 | -0.31** | 2.77 | 2.76 | 1.57 | -0.55 | -2.09 |
| 3     | 0.31** | 0.34*** | 0.16* | 0.01 | -0.27** | 2.09 | 2.87 | 1.82 | 0.08 | -2.16 |
| 4     | 0.31*** | 0.14 | -0.01 | -0.04 | -0.44*** | 2.82 | 1.20 | -0.07 | -0.29 | -2.82 |
| Big   | 0.64*** | 0.71*** | 0.31** | -0.07 | -0.69*** | 2.88 | 5.87 | 2.49 | -0.69 | -8.18 |

It can be seen from table 4.8 that as the B/M of the portfolios become higher, the HML coefficient gradually increases, and in the fifth group with the highest B/M ratio, the coefficient is significantly different from zero, indicating that the A-share market has a significant market value effect. In addition, the coefficients of HML in the table are significant in many places, indicating that HML factors have very strong explanatory ability; The table 4.9 shows that the factor coefficient of the portfolio with conservative investment style is significantly positive, while the factor coefficient of the portfolio with radical investment style is significantly negative, which also shows that the investment effect of the A-share market is obvious. And its CMA coefficient is also significant in many points, indicating its explanatory ability is still good.

Although CMA and HML factors are not significant in Picture 4.3, these two factors still have strong explanatory ability, which also supports the conclusions of GRS test below. Our guess is that the time span is long. China's stock market is greatly affected by different policies at different stages, and the investment style of investors has also changed, diluting the value effect and investment effect. Therefore, there is two redundant factors. We will give a detailed explanation below.

**(4) Comparison of factor models based on GRS test**

(4.1) Introduction of GRS test

GRS test was proposed by Gibbons, Ross and Shaken in 1989, and also used in Fama-French five-factor model which is a quantitative test method for the validity of the factor model. The null hypothesis of this test is that the intercept term of the factor model is equal to zero: $a_i=0$. The statistics of GRS submit the F distribution with degrees of freedom n and t-n-l, n represents the number of portfolios participating in the regression, l represents the number of factors in the model, and t represents the number of observation periods of data.

It should be noted that because there are too many factors affecting the excess returns of portfolio in the stock market, the asset pricing models including the three-factor model and the five-factor model all cannot fully explain the change of the return rate of portfolio. Therefore, the purpose of this GRS test is not to test whether the three-factor and the five-factor models can reject the null, but to compare the explanatory power (validity) of the two models in China's A-share market based on the GRS statistics from the two models.

(4.2) Results of GRS test

We select the monthly data of China's A-share market from 2005 to 2020 and use Stata's GRS function(grstest2) for regression and test. In order to evaluate the performance of the two models in fitting the average monthly return of the 5 * 5 portfolio grouped by Size-B/M, Size-OP and Size-Inv, we calculated the GRS statistics of both the three-factor and the five-factor models in the above three cases and the mean of the absolute value of intercept term obtained by regression($A|a_i|$). The results of GRS test are as follows:

Table 4.10 GRS test results of different factor models

| 25 Size-BM | GRS | $A|a_i|$ |
|---|---|---|
| MKT SMB HML | 2.508*** | 0.019 |
| MKT SMB HML RMW CMA | 2.389*** | 0.101 |
| 25 Size-OP | GRS | $A|a_i|$ |
| MKT SMB HML | 3.073*** | 0.090 |
| MKT SMB HML RMW CMA | 2.377*** | 0.049 |
| 25 Size-Inv | GRS | $A|a_i|$ |
| MKT SMB HML | 2.226*** | 0.048 |
| MKT SMB HML RMW CMA | 1.670*** | 0.067 |

It can be seen from table 4.10 that the intercept values of the two models tend to zero at the significant level of 0.05, which means the three-factor and the five-factor models are both competent to explain the excess return of portfolio. Under the portfolio of Size-

B/M, Size-OP and Size-Inv, the GRS statistics of the five-factor model are smaller than those of the three-factor model, which indicates that the explanation ability(validity) of the five-factor model is better than that of the three-factor model in China's stock market.

It can be seen from tables 4-2, 4-3 and 4-4 that although the profitability and investment level of Chinese companies have no significant impact on the excess returns of their stocks, the average return of the stock market still shows certain characteristics of profitability effect and investment effect. Therefore, the reason why Fama-French's five-factor model has better explanatory power is that it introduces these presented effects into the model in the form of factors. Fama-French's five-factor model, due to the influence of market, size, B/M, profitability and investment level on the asset prices of China's stock market, can better explain the characteristics of these effects existing in the average return of China's stock market.

In addition, interestingly, the results of our paper are obviously different from those of some scholars. For example, Zhao Shengmin (2016), using China's stock data from January 1995 to December 2014, concluded that the explanatory power of the three-factor model is better than that of the five-factor model.

The possible explanations for this difference are as follows:
1. Different stock selection strategies (data processing method): in our paper, except for ST and PT, we also exclude the data of the first six months after IPO (including the listing month) and exclude the data of financial stocks. Different stock selection strategies may have a certain impact on the conclusion.

2. Changes in China's stock market policy: Shanghai and Shenzhen stock exchange of China were established in late 1990. Some of the research used old stock data, considering the situation of immature period of Chinese stock market, which may change greatly with the current situation; affected by the policy (for example, China implemented the 'Price Limit' at the end of 1996), the conclusions may be different. In addition, in the early stage of China's stock market, the shares of listed companies in the A-share market were divided into tradable shares and non-tradable shares. About two-thirds of the shares could not be traded and were held by the government or state-owned companies, which is a unique feature of China's A-share market. Chinese government began the share-structure reform in 2005 and legally converted non-tradable shares into fully tradable shares at the end of that year. Therefore, this study selects the data after 2005 to ensure that the nature of the stock market is as similar as possible with other regions' such as the United States. From the perspective of the company or industry, before reform, the investment efficiency of most state-owned listed companies was low, and capital abuse and over investment could be often seen. The share-structure reform is conducive to the introduction of market-oriented incentive mechanism, further improve the company's investment confidence and strengthen profitability. Li Zhibing (2017) tested the five-factor model in different periods and believed that after the reform, the A-share market showed similar investment and profit effects to the U.S. stock market, that is, the five-factor model was more applicable. This also echoes the phenomena in tables 4.3 and 4.4 above, making the characteristic effects related to profit and investment more obvious.

3. Under the standardized information disclosure rules, the annual financial statements of listed companies are generally published in March and April of the following year, which means the end of April is the deadline for the disclosure of annual reports. The lag of information disclosure of financial statements results in the inconsistency between financial data in financial database and market data. Some studies may adopt the method of constructing the portfolio at the end of December in the previous year, and such method used when the accounting annual report is not published may have an impact on the results.

## 4. Conclusion summary and analysis

In this paper through regression coefficient test and GRS test, taking the companies' financial data and stock market data of China's A-share market from 2005 to 2020 years as the original data, we analyze whether Fama-French's five-factor model has a high enough explanatory power to the change of excess returns of China's A-share market with the change of time.

The results of this empirical analysis show that:

(1) By referring to the empirical test method of "5x5" proposed by Fama French, we find that the change of excess return of portfolio in China's A-share market witness a very obvious Size effect, which indicates the company portfolio with high market value has a higher stock excess return than the company portfolio with low market value. At the same time, China's A-share market also has obvious value effect. Stocks with high B/M tend to enjoy higher excess returns. In addition, there is no significant relationship between the profitability and investment effect of the portfolio and the stock excess return of their groups, but there is still a certain characteristic effect related to the profitability, investment and excess return.

(2) In the part of regression coefficient test, we use STATA to do the OLS regression applying Fama-French's five-factor model to the stock data of China's A-share market from January 2005 to December 20. We find that the coefficients of factors MKT, SMB and RMW are obviously different from zero, but the HML and the CMA are not. However, after 5*5 portfolio regression analysis, these two factors still have strong explanatory ability. This is different from the empirical result of Fama-French (2015) in the US stock market (It is considered that only the HML is redundant variable compared with RMW and CMA).

(3) The result of GRS test show that both the three-factor model and the five-factor model can accept the null hypothesis of GRS test at the significance level of 5%, and the five-factor model has better explanatory power(validity) than the three-factor model.

To sum up, the empirical test shows that: in China's A-share market, the HML and CMA are redundant but still with explanatory ability, while the remaining SMB, MKT and RMW have very obvious explanatory effect on the excess return of A-share market portfolio. Therefore, we believe that although the explanatory power of the five-factor model in China's A-share market is not as good as that in the US market, the five-factor model still shows better explanatory power, compared with the three-factor model.

The differences of explanatory power (validity) between the three-factor model and the five-factor model can reflect the differences in asset selecting and investment decision-making between the two markets. In the past, China's investment market is mainly dominated by individual investors. There are obvious opportunistic effect and herding effect in investment decision-making. The understanding of company value and investment status are not the main influencing factors when those investors make investment decisions. However, under the condition that the rationality and investment level of each trader have been improved, the investment preference of each trader has changed, and the investment strategy of "buying the winners and selling the losers" has been criticized by more and more investors. Investors pay more and more attention to the development prospects and growth space of listed companies and industries. In recent years, more and more investors in China's investment market have begun to choose to hand over their own money to professional institutions. Private placement, ETF, FoF and other fund investments are more popular. The investment market shows the characteristics of "de-retail", which is also the reason why CMA investment factors have strong explanatory power in the A-share market. This trend will be more obvious in the future.

In contrast, in the mature investment market, institutional investors in the United States have obvious advantages. Their professional investment ability, technology and performance have been recognized by the market for many years. Moreover, financial products are rich and diverse with relatively reasonable handling fees, which has been trusted by the majority of individual investors in the United States. Its investment style is significant, and the phenomenon that the CMA is redundant would not happen.

The turnover rate of China's stock market in recent years has also decreased between 200% and 300%, compared with that of more than 500% in earlier years. It also shows that some investors have changed their investment strategies. Instead of investing for the purpose of getting income from the short-term stock price fluctuation, they choose to hold, pay more attention to the company's financial information and investment decisions, and make long-term investment. However, the turnover rate of US stocks is only about 100%. This is also the reason why Fama-French five-factor model enjoy better explanatory power in the U.S. market. Moreover, the Chinese market is greatly affected by the policies, which will weaken the company's profitability and investment effect.

In addition, when we only add CMA to the GRS test of the four-factor model, the GRS statistics of the three portfolios are still lower than the three-factor one, but higher than the five-factor GRS statistics, which shows that CMA factor has the explanation ability.

B/M is the reciprocal of P/B ratio. The HML represents the valuation and investment value of the company, while RMW and CMA reflect the growth possibility of the company. We speculate that RMW and CMA may contain some information of HML, which makes HML redundant after adding two factors. We calculated the correlation coefficient matrix of the five factors and found that the correlation coefficient between HML factor and RMW factor was 0.000637, while the correlation coefficient with CMA factor was higher, 0.475 and being significant, indicating that CMA factor contained some information of HML, but there was still no adverse effect caused by collinearity.

In the Chinese market, most stock trading platforms will provide users with P/B ratio data. This indicator is relatively intuitive. Compared with the investment situation, development prospects and other data of companies, the P/B ratio is the data that individual investors can directly obtain. Therefore, stocks with undervalued value are often selected for investment according to this indicator. Moreover, investment institutions often take the P/B ratio as one of the reference indexes, and often use this index to estimate company's market value. The HML has a great relationship with P/B ratio, which is also the reason why HML factor has a strong ability to explain stock returns.

From those perspectives, we could also try to put forward the expectation and development suggestions for investment field and relative research about asset pricing:

First of all, the effectiveness and explanatory power of asset pricing model depend on the situation of the investment market. In order to put forward a suitable factor model which can more appropriately explain the change of returns, the construction of asset pricing model needs to fully consider the actual situation of the investment market and the environment in which it is located. Secondly, China's A-share market is still in the stage of development. There are some problems such as the low degree of market regulation, information opacity and asymmetry. Also, the market information disclosure mechanism still needs improvement. There are also some phenomena such as false disclosure of company information and insider's use of false information for stock operation, which may lead to investors' distrust to the company financial information. Therefore, the continuous development of the field of asset pricing is inseparable from the scholars' in-depth research and repeated empirical tests on the asset pricing model, and also inseparable from the market's sound regulatory mechanism to ensure the market's stability and development.

**Acknowledgments:** We acknowledge funding from the National Natural Science Foundation of China (Nos. 72071051, 71871071) and the Key Program of the National Social Science Foundation of China (No. 21AZD071).